\documentclass[reprint, aps,superscriptaddress,amsmath,amssym,prb]{revtex4-2}

\usepackage{bm}
\usepackage{float}
\usepackage{graphicx}
\usepackage[usenames,dvipsnames]{color}
\usepackage[normalem]{ulem}
\usepackage{bm}
\usepackage{multirow}
\usepackage{titlesec}
\usepackage{ragged2e}
\usepackage[english]{babel}
\usepackage[T1]{fontenc}
\usepackage[table]{xcolor}
\usepackage[utf8]{inputenc}
\usepackage{lmodern}
\usepackage{babel}
\usepackage{amsmath}
\usepackage{amssymb}
\usepackage{envmath}
\usepackage{makecell}
\usepackage{comment}
\usepackage{geometry}
\usepackage{caption}
\usepackage{subcaption}
\usepackage{stackengine}
\usepackage{graphicx}
\usepackage{tikz}
\graphicspath{{images/}}
\usepackage{array}
\newcolumntype{P}[1]{>{\centering\arraybackslash}p{#1}}
\usepackage{parskip}
\usepackage[font=small,labelfont=bf,
   justification=justified,
   format=plain]{caption}
\usepackage{wrapfig}
\begin{document}
\title{Electron-electron and electron-phonon  collision cross sections in CsV$_3$Sb$_5$}

\author{Charles Menil}
\affiliation{Laboratoire de Physique et d'\'Etude de Mat\'{e}riaux (CNRS)\\ ESPCI Paris, PSL Research University, 75005 Paris, France }
\author{Andrea Capa Salinas}
\affiliation{Materials Department, University of California\\ Santa Barbara, California 93106, USA}
\author{Stephen D. Wilson}
\affiliation{Materials Department, University of California\\ Santa Barbara, California 93106, USA}
\author{Beno\^it Fauqu\'e}
\affiliation{Laboratoire de Physique et d'\'Etude de Mat\'{e}riaux (CNRS)\\ ESPCI Paris, PSL Research University, 75005 Paris, France }
\author{Kamran Behnia}
\affiliation{Laboratoire de Physique et d'\'Etude de Mat\'{e}riaux (CNRS)\\ ESPCI Paris, PSL Research University, 75005 Paris, France }

\date{\today}

\begin{abstract}
AV$_3$Sb$_5$ (A = K, Rb, Cs) are kagome metals and superconductors, attracting much recent attention as nexus of multiple quantum states. Here, through a systematic study of electric and thermal transport of CsV$_3$Sb$_5$, we identify it as a metallic Fermi liquid with moderate electronic correlations and strong electron–phonon (e-ph) collision cross section.  We observe contributions to the inelastic electrical resistivity, each dominating within a distinct temperature window. The prefactor of the T$^2$ is consistent with the Kadowaki-Woods scaling for a Fermi liquid with moderate correlation. By performing thermal conductivity measurements at zero and finite magnetic field, we separate the electronic and the lattice contributions to the thermal conductivity. The Wiedemann-Franz law is satisfied in the zero-temperature limit, while a downward deviation emerges at finite temperature due to the mismatch between the prefactors of the electrical and thermal quadratic resistivities, as reported in other metals. The Bloch–Grüneisen description of electron–phonon scattering successfully accounts for both electronic thermal and electrical transport, indicating a remarkably large e–ph collision cross section in CsV$_3$Sb$_5$.
\end{abstract}
\maketitle

\section{Introduction}
Discovered in 2019, the AV$_3$Sb$_5$ (A=K, Rb, Cs) family are kagome metals with a  Charge Density Wave (CDW) instability and a superconducting ground state \cite{ortiz2019new,ortiz2020cs}. They have been a subject of much attention as a platform to explore the interplay between geometric frustration and  collective electronic phenomena. 

The rich phase diagram of  CsV$_3$Sb$_5$ has been experimentally mapped \cite{wilson2024v3sb5}. The possible loss of time reversal symmetry in the CDW state despite the absence of any magnetic order \cite{xu2022three,khasanov2022time,gui2025probing,saykin2023high,guo2024correlated,jiang2023kagome,wilson2024v3sb5,neupert2022charge, park2021electronic,liege2024search} has been one debated topic. In addition to the CDW, with an onset of $\approx$ 100 K (the onset of CDW), and  superconductivity, with a critical temperature of $\approx$ 3 K, other anomalies have been reported, notably at  $\approx$ 30 K \cite{chen2022anomalous,wei2024three} and $\approx$ 60 K \cite{stahl2022temperature} and a nematic state has also been invoked \cite{nie2022charge}.

Thanks to studies of quantum oscillations \cite{Ortiz2021,Huang2022,Shrestha,Broyles,Zhang2022,Zhang2024}, the Fermi surface of this metal is fairly well documented. The reconstruction of the Fermi surface by the CDW  \cite{Ortiz2021},  generates numerous detectable frequencies. Despite its layered structure, this metal hosts three-dimensional pockets \cite{Huang2022}. Pressure destroys the CDW and the Fermi surface reconstruction. A study of quantum oscillations under pressure has explored the non-reconstructed Fermi surface  \cite{Zhang2024} and found it in agreement with DFT expectations. 

Despite the wealth of information gathered by these studies, a number of basic questions have not been explicitly answered. Does the electronic transport in this metal correspond to what is expected in a Fermi liquid? If yes, what is the amplitude of electron-electron scattering and how does it compare to other metallic Fermi liquids? What about electron-phonon scattering? What is the share of electronic and phononic contributions to the total thermal conductivity? Is there any sign of electron-electron scattering in thermal channel?  What is the effect of phonon scattering on the electronic thermal conductivity compared to other metals?

Following  previous studies of transport in CsV$_3$Sb$_5$ \cite{pang2023glasslike,kountz2024thermal, chen2022anomalous,wei2024three,hossain2025unconventional}, we report on a study electric and thermal transport quantifying the amplitude of the e-e and e-ph scattering contributions to the electric and thermal resistivities for the first time. According to our findings, CsV$_3$Sb$_5$ is a Fermi liquid with electronic correlations of moderate strength. On the other hand, it has a remarkably large electron-phonon collision cross section, presumably due to the vicinity of a structural transition. We quantify the respective share of scattering by electrons and phonons to the electric conductivity and to the electronic component of the thermal conductivity. We also separates the electronic and the phononic contributions to heat transport. The former obeys the Wiedemann-Franz law in the zero-temperature limit but deviates downward at finite temperature. The latter implies a phonon mean-free-path well below the sample size. Thus, according to our study, the standard picture of solid-state transport gives a reasonable account of  in-plane electric and thermal resistivity in this solid. It provides quantitative numbers to be confronted with what, at least in principle, can be computed from the specific electronic and structural properties of this metal. 

\begin{figure}[hbt!]
    \centering
     \begin{subfigure}{0.4\textwidth}
        \captionsetup{justification=raggedright,singlelinecheck=false}
        \caption{}
        \includegraphics[width=1\textwidth]{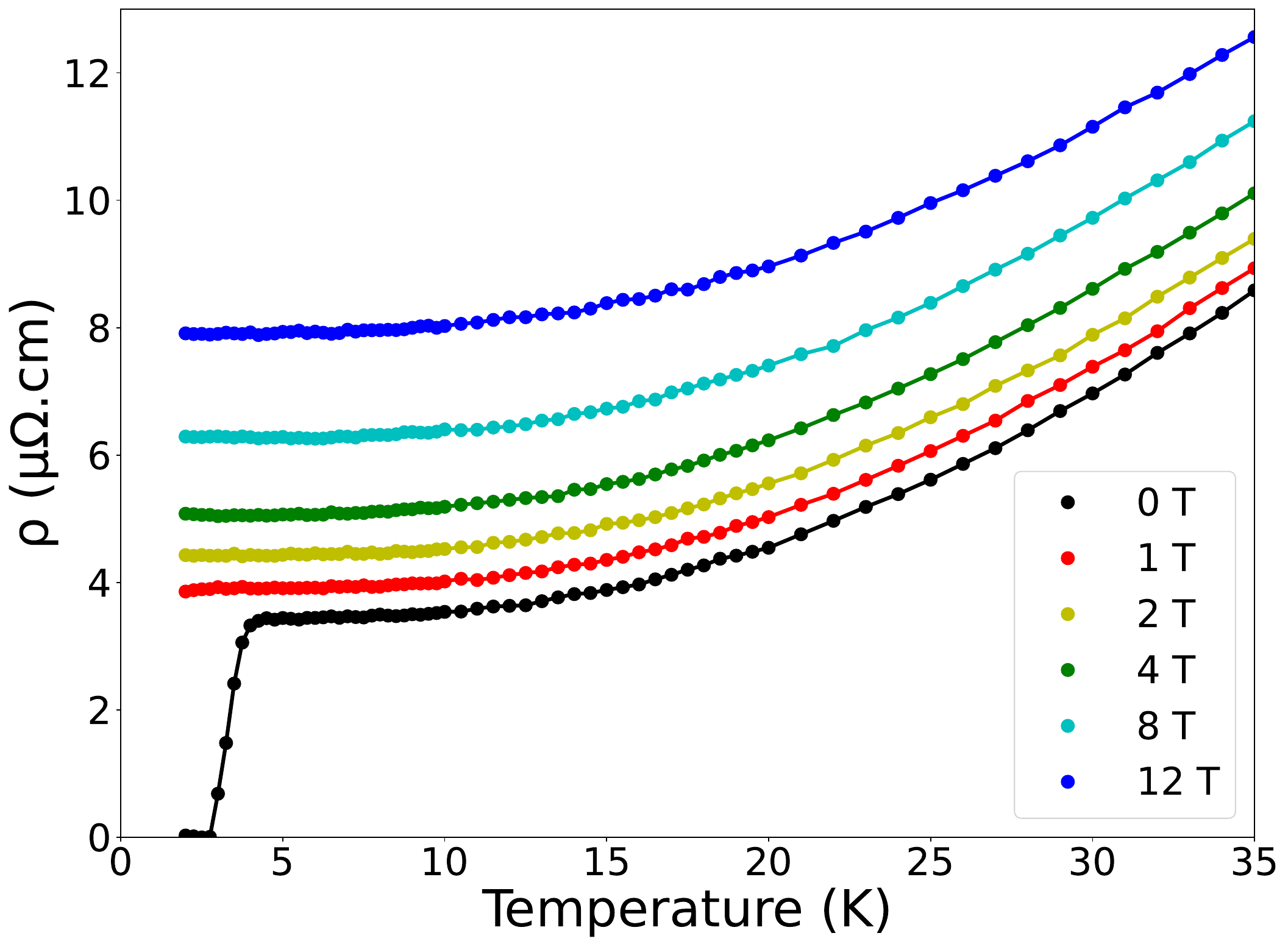}
     \end{subfigure}
     \begin{subfigure}{0.4\textwidth}
        \captionsetup{justification=raggedright,singlelinecheck=false}
        \caption{}
        \includegraphics[width=1\textwidth]{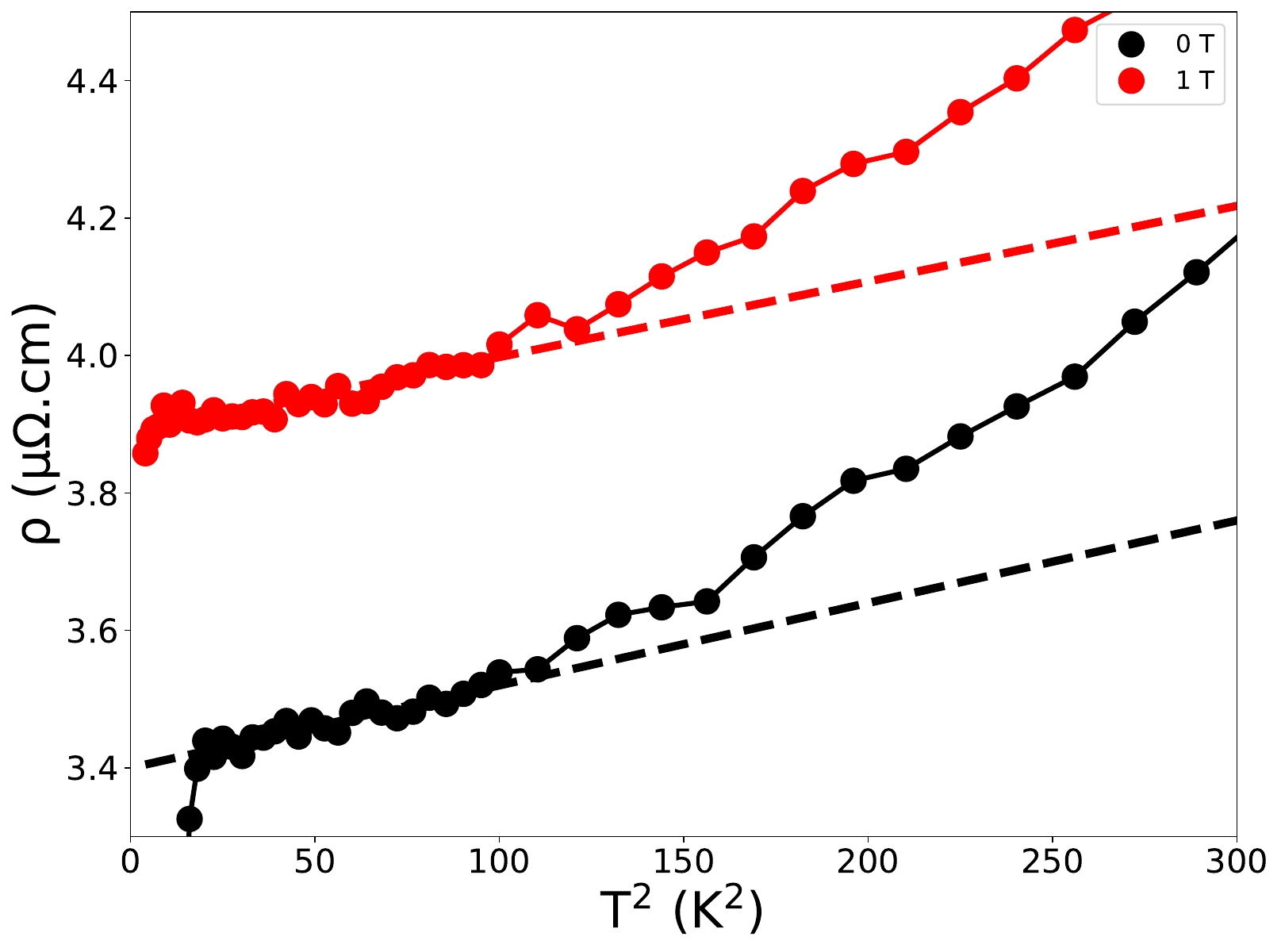}
     \end{subfigure}
     \begin{subfigure}{0.4\textwidth}
        \captionsetup{justification=raggedright,singlelinecheck=false}
        \caption{}
        \includegraphics[width=1\textwidth]{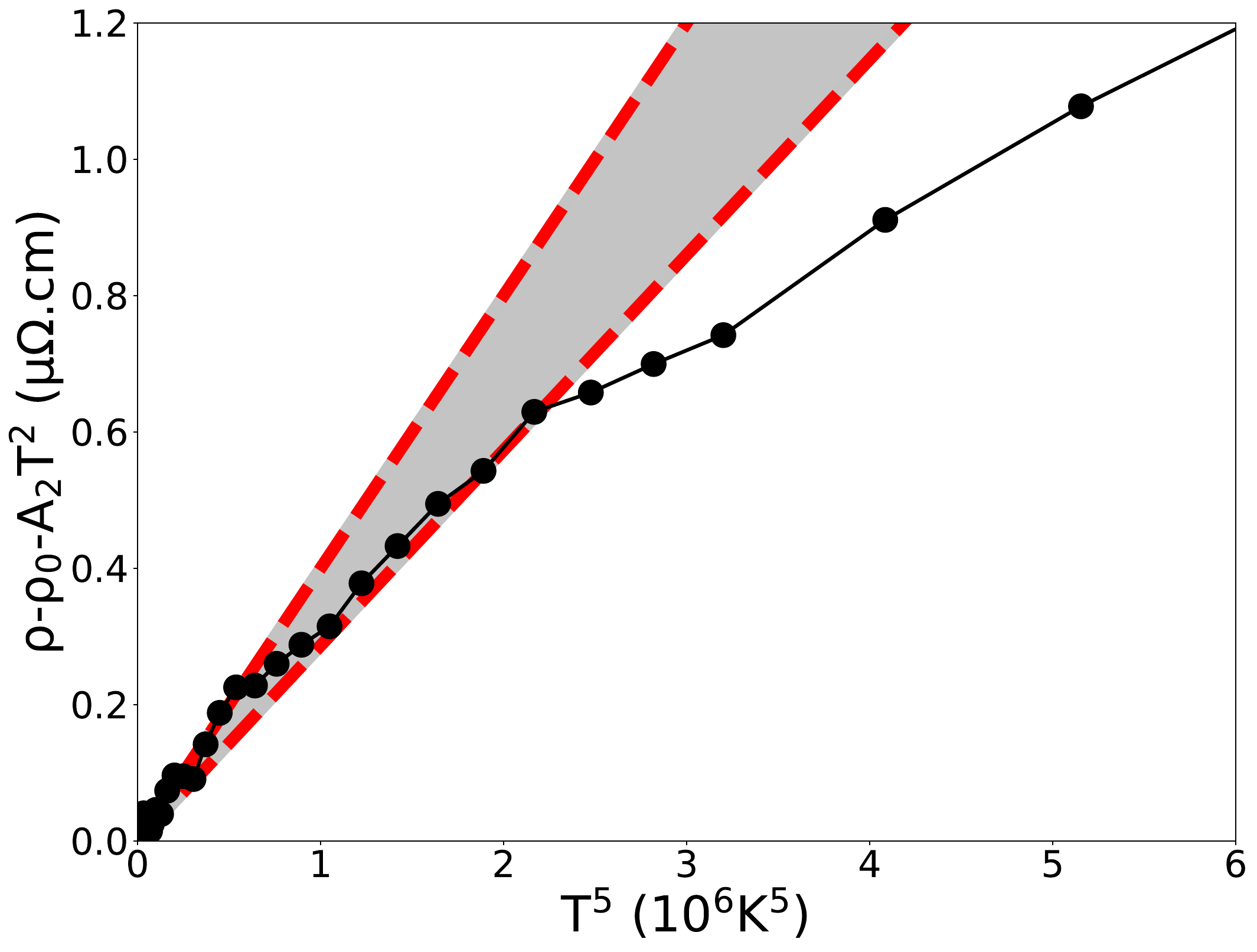}
     \end{subfigure}
     \caption{\justifying{Low Temperatures electrical resistivity of CsV$_3$Sb$_5$. (a) Resistivity from 2 K to 35 K for several applied magnetic fields along crystal c-axis. (b) Resistivity at zero magnetic field and 1 T versus temperature square. Dash lines show T-square resistivity for temperatures between 4.5 K and 10 K. (c) Residual resistivity at zero magnetic field versus temperature to the power 5. The red lines show estimated minimum and maximum T$^5$ resistivity. The gray area shows the incertitude on the T$^5$ resistivity.}}
    \label{fig: Rho}
\end{figure}

\begin{figure}[ht]
    \centering
     \begin{subfigure}{0.4\textwidth}
        \captionsetup{justification=raggedright,singlelinecheck=false}
        \caption{}
        \includegraphics[width=1\textwidth]{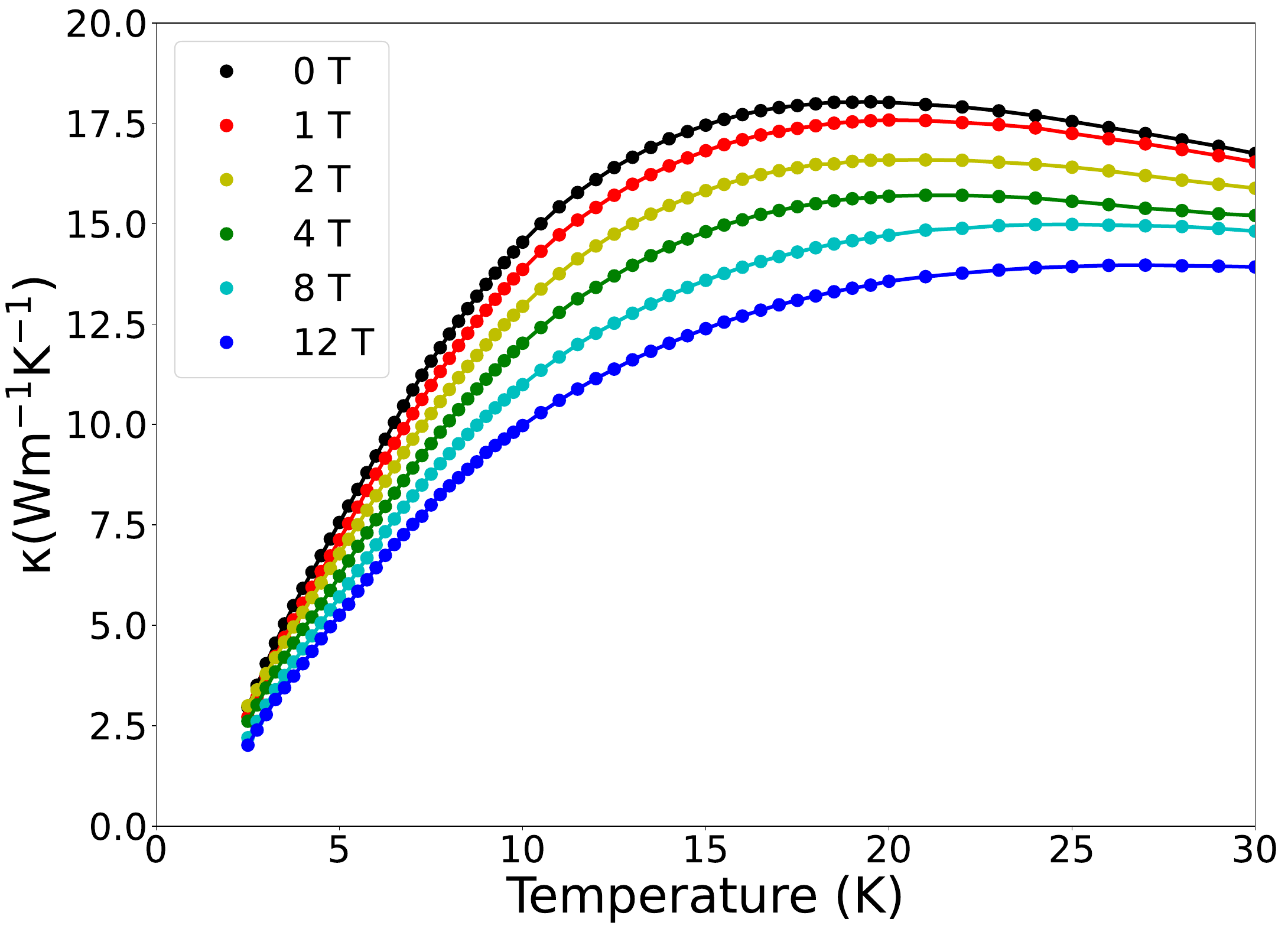}
     \end{subfigure}
     \begin{subfigure}{0.4\textwidth}
        \captionsetup{justification=raggedright,singlelinecheck=false}
        \caption{}
        \includegraphics[width=1\textwidth]{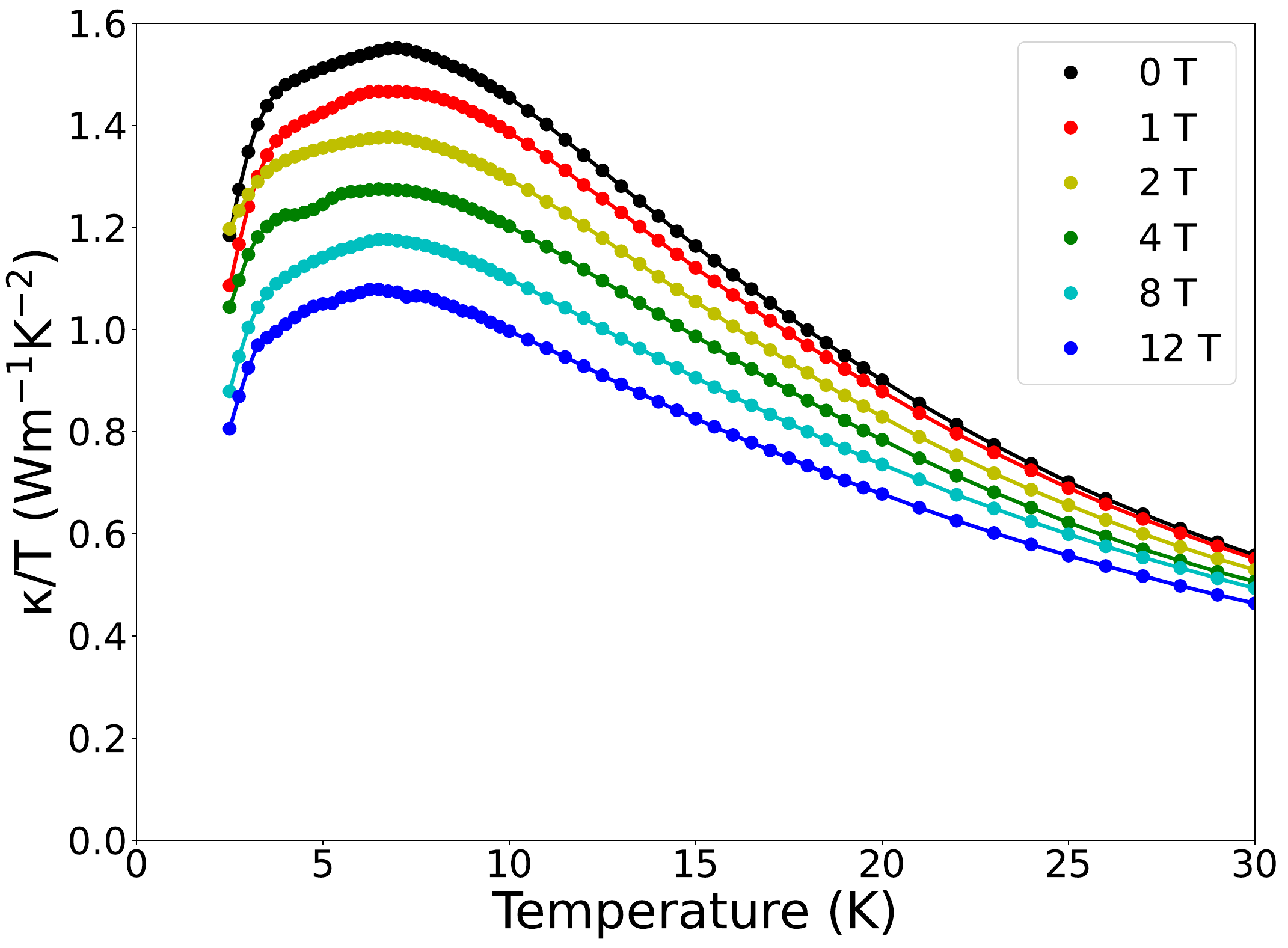}
     \end{subfigure}
     \caption{\justifying{Thermal conductivity of CsV$_3$Sb$_5$ crystal. (a) Thermal conductivity from 2 K to 30 K for several applied magnetic fields along crystal c-axis. (b) Thermal conductivity divided by temperature, from 2 K to 30 K. Resistivity and thermal conductivity have been measured on the same sample. }}
    \label{fig: Kappa}
\end{figure}

\begin{figure*}[ht]
    \centering
     \begin{subfigure}{0.4\textwidth}
        \captionsetup{justification=raggedright,singlelinecheck=false}
        \caption{}
        \includegraphics[width=1\textwidth]{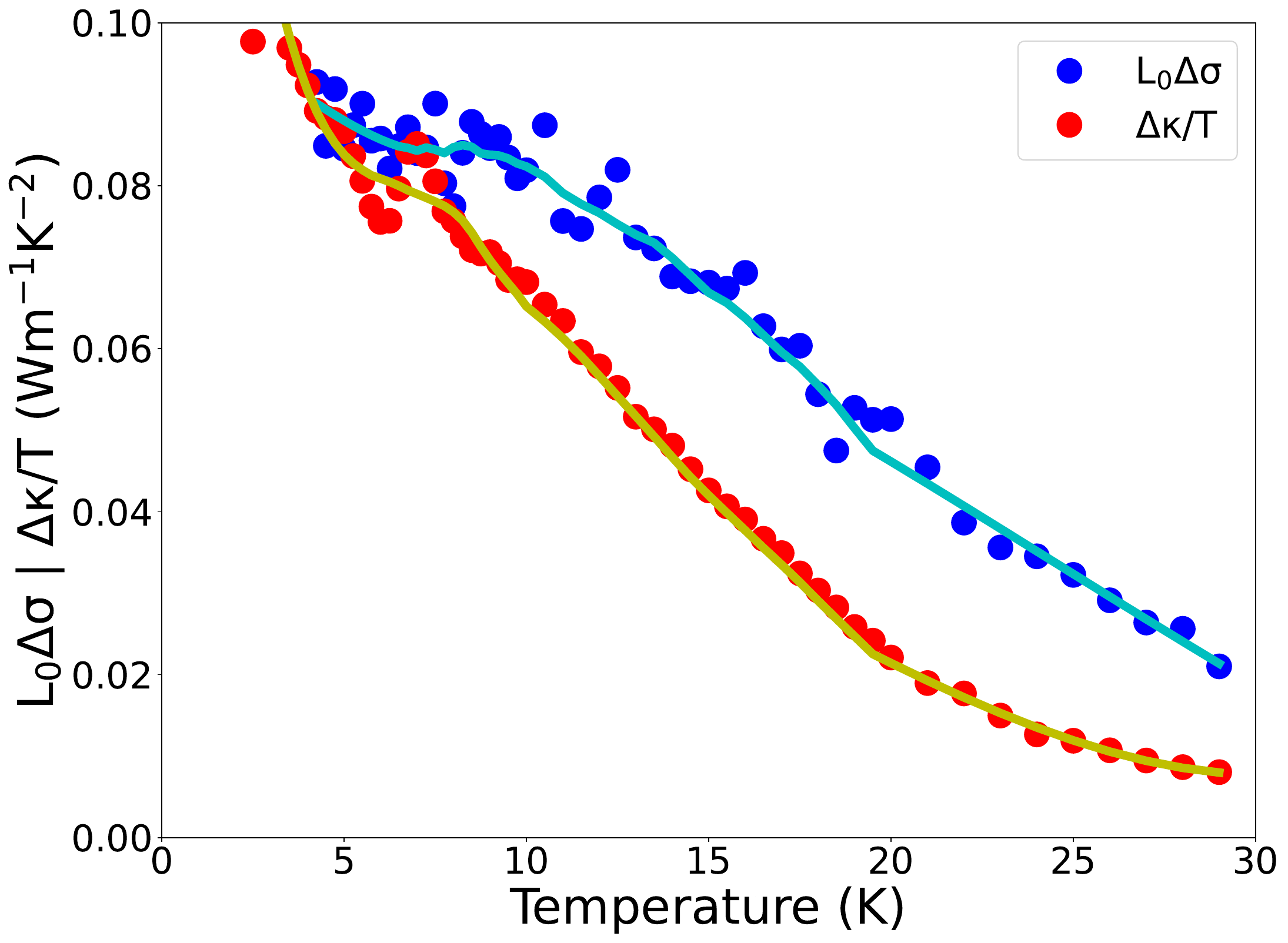}
     \end{subfigure}
     \begin{subfigure}{0.4\textwidth}
        \captionsetup{justification=raggedright,singlelinecheck=false}
        \caption{}
        \includegraphics[width=1\textwidth]{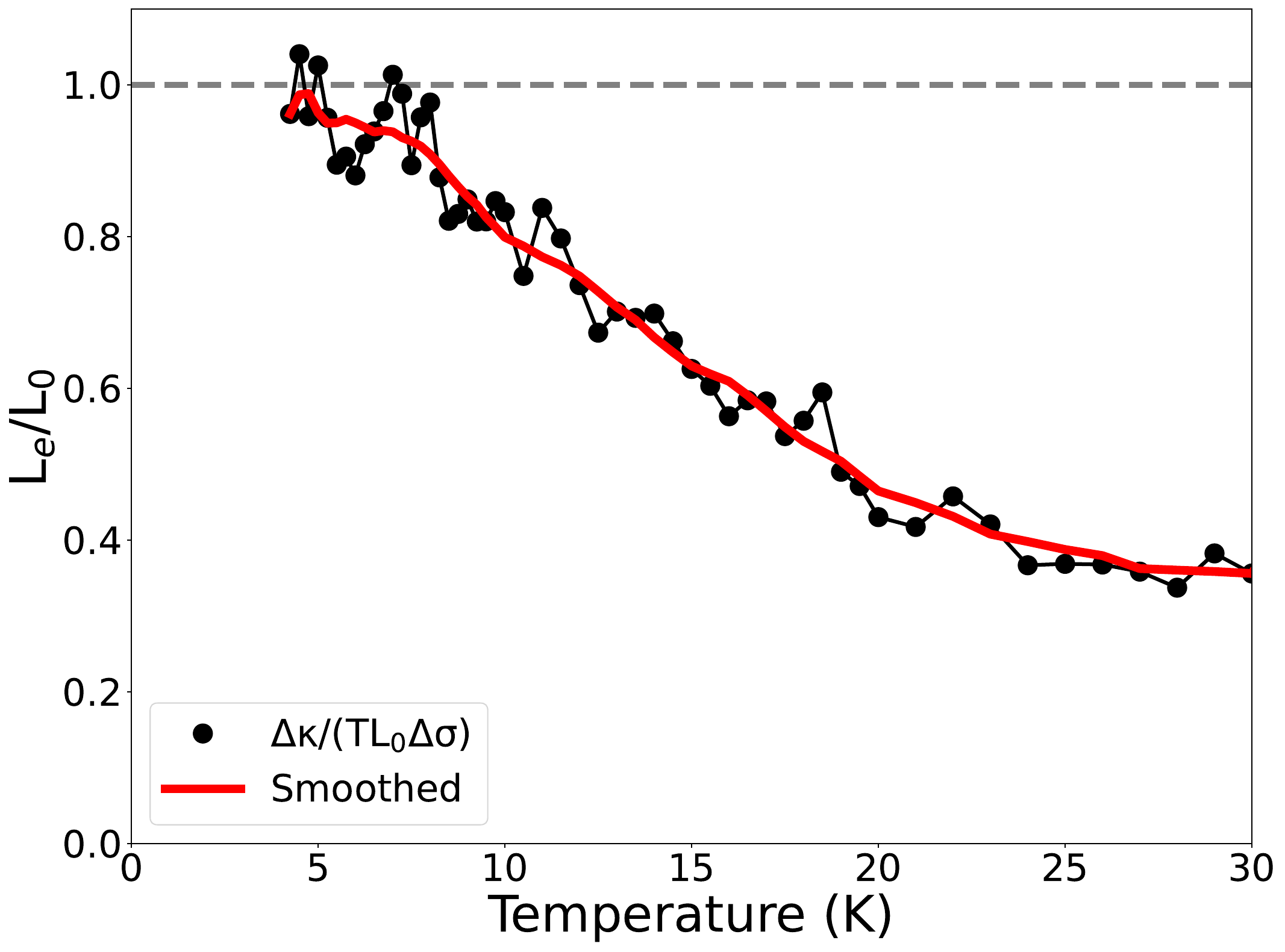}
     \end{subfigure}
    \begin{subfigure}[b]{0.4\textwidth}
         \centering
          \captionsetup{justification=raggedright,singlelinecheck=false}
         \caption{}
      \includegraphics[width=\textwidth]{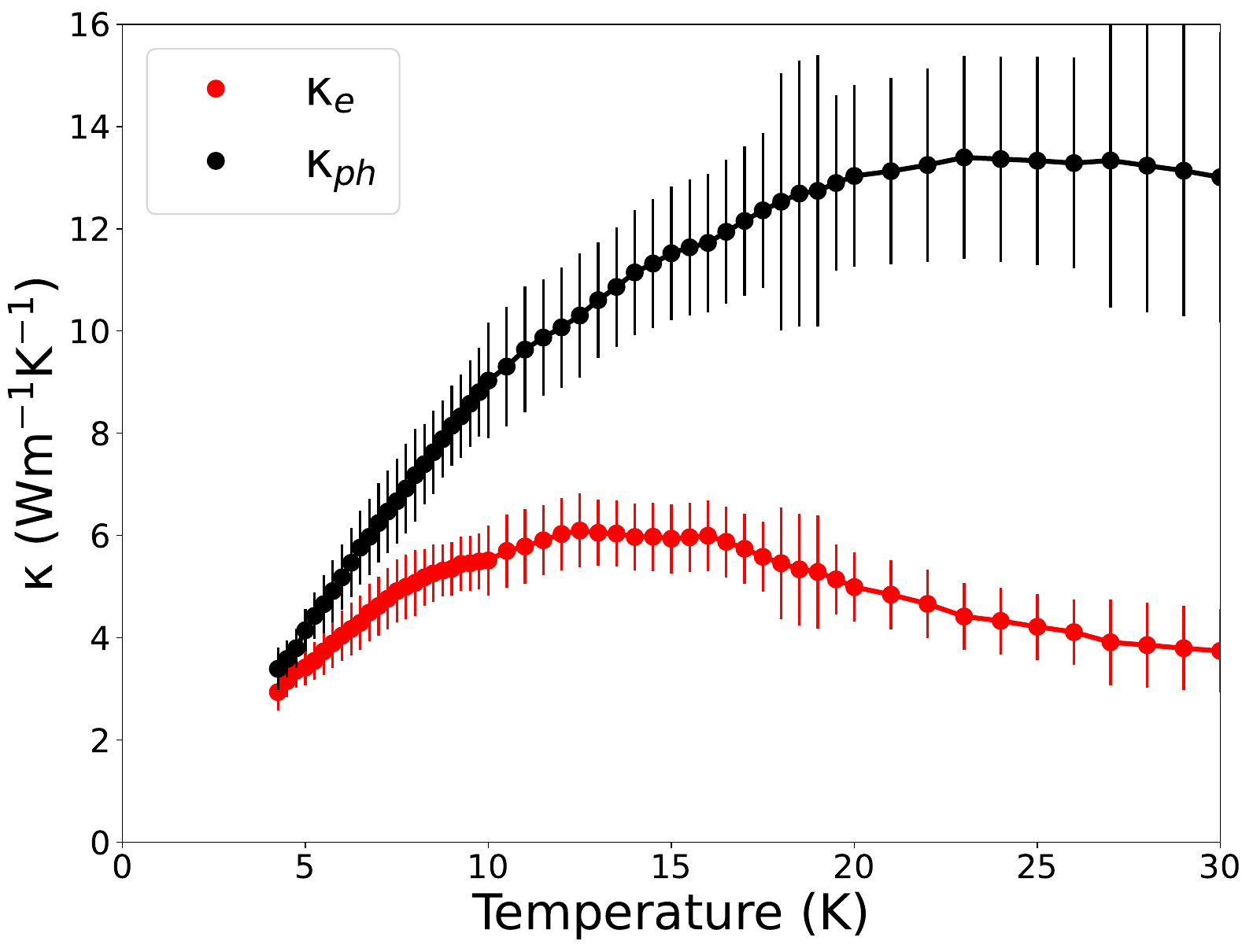}
     \end{subfigure}
     \begin{subfigure}[b]{0.4\textwidth}
         \centering
          \captionsetup{justification=raggedright,singlelinecheck=false}
         \caption{}
      \includegraphics[width=\textwidth]{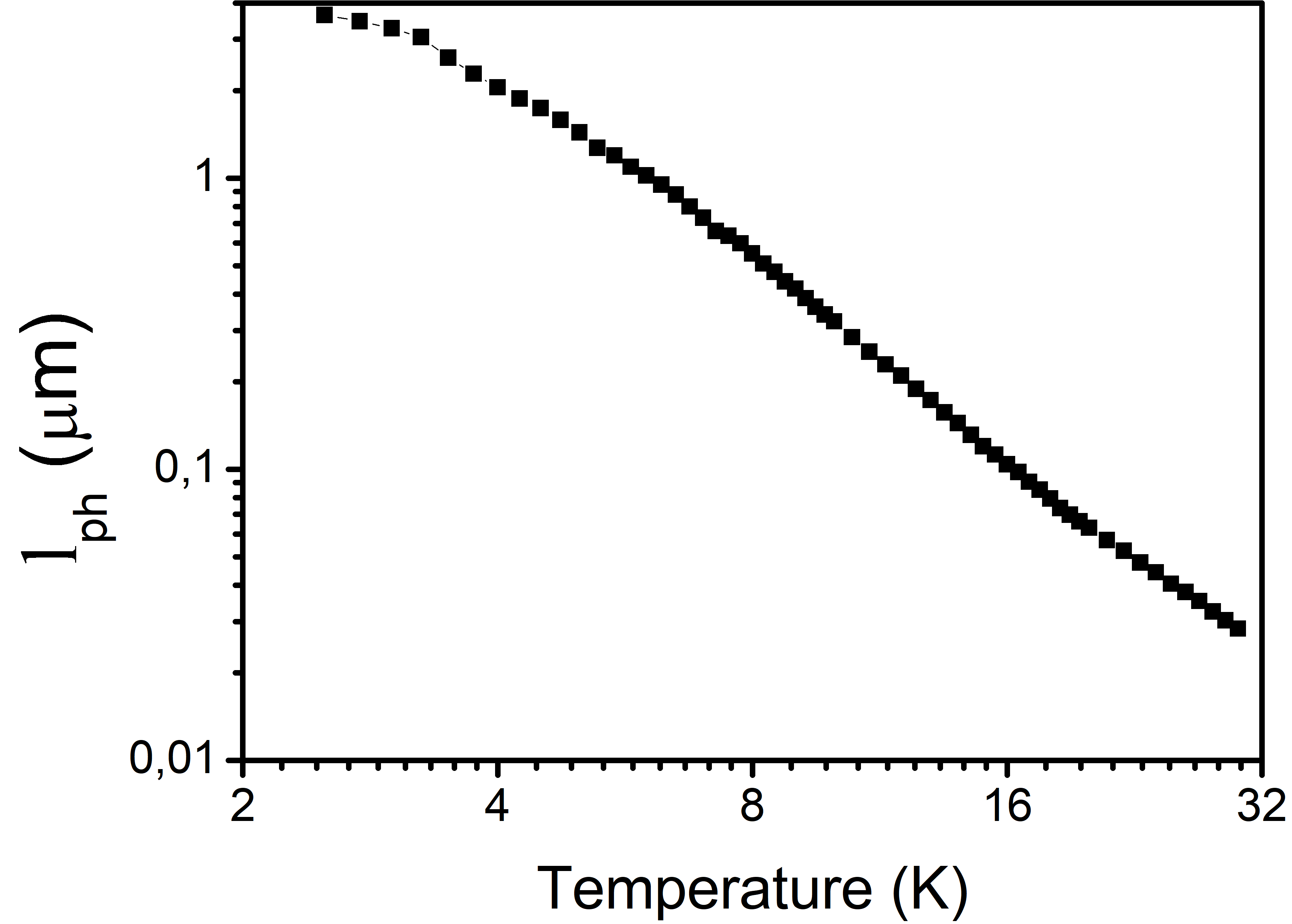}
     \end{subfigure}
     \caption{\justifying{Separation between electrons and phonon contribution in thermal conductivity. (a) Difference in electronic and thermal conductivity between measurements under 0 T and 1 T. (b) L divided by L$_0$ obtained from measurements at 0 T and 1 T. (c) Phononic and electronic thermal conductivity under zero magnetic field. (d) The phonon mean free path, l$_{ph}$ {\it{vs}}. temperature, in a log-log scale. It has been calculated using the kinetic formula $\kappa=\frac{1}{3}C v_s l_{ph}$, the reported sound velocity ($v_s=$1960 m.s$^{-1}$ \cite{pang2023glasslike}) and the experimentally measured heat capacity \cite{yang2023charge}.}}
    \label{fig: separation}
\end{figure*}
\section{Results}
The in-plane electrical resistivity, $\rho$, of a CsV$_3$Sb$_5$  single crystal measured by a four-electrode method, is shown in Figure \ref{fig: Rho}a. At zero-field, the superconducting transition is visible at $\approx$ 3 K. Upon the application of a magnetic field of 1 T, perpendicular to the kagome planes, the superconducting transition disappears. Resistivity  curves shift vertically and almost rigidly with magnetic field. 

Figure \ref{fig: Rho}b is a plot of zero-field resistivity as function of the square of temperature. One can see that between the superconducting transition and  $\approx$ 10 K, $\rho$ is linear in $T^2$. A linear fit in this narrow window yields a finite intercept at T=0 : $\rho_0$. At high temperature, the data deviates upward from this linear fit. This deviation signals the presence of a larger exponent for inelastic resistivity at higher temperatures. The most plausible candidate is a T$^5$ dependence, expected by the Bloch-Gr\"uneisen law of electron-phonon scattering when $T \ll\Theta_D$. This hypothesis is confirmed in Fig. \ref{fig: Rho}c. It is a plot of resistivity, after the subtraction of its constant ($\rho_0$) and quadratic ($A_2T^2$) terms,  as function of $T^5$. Below 18 K, one can detect a $T^5$, with an uncertainty on the slope, which is highlighted by the shaded area. Note the downward deviation at higher temperatures, signaling evolution towards a lower exponent. This conforms to  the Bloch-Gr\"uneisen picture, where the exponent is expected to gradually evolve towards 1, as the Debye temperature ($\Theta_D \approx 160$ K \cite{yang2023charge}) is approached.   

Thus, we conclude that below 18 K (that is, $\approx\Theta_D/10$), $\rho$, follows:

\begin{equation}
    \rho=\rho_0+A_2T^2+A_5T^5
\end{equation}

Here, $\rho_0$ is the residual resistivity, $A_2$ is the $T$-square prefactor, caused by e-e scattering and $A_5$, the $T^5$ prefactor, due to e-ph scattering. 

A fit to our zero-field data yields $\rho_0= 3.4$ $\mu \Omega$.cm, $A_2=1.2\pm0.2$ n$\Omega$.cm.K$^{-2}$ and $A_5=2.6\pm0.2$ $\times 10^{-7}\mu\Omega$.cm.K$^{-5}$. In contrast to the residual resistivity, which is extrinsic and is set by the concentration of defects and impurities, the two other terms are intrinsic to the metal in question.  Mott \cite{mott2004metal} argued that $A_2$ is proportional to the collision cross section between two electrons regardless of the microscopic mechanism leading to dissipation \cite{lin2015}. We will use the same expression for the size of the phase space of the scattering between electrons and phonons.

The quadratic term, $AT^2$, is ubiquitous in metals, save for exceptions, such as 'strange' \cite{Legros2019} or 'polar' \cite{Wang2019} ones. It is  caused by electron-electron scattering \cite{Rice1968,KADOWAKI,behnia2022origin}. Since the phase space of such collisions is proportional to $(k_BT/E_F)^2$, it is easily observable in dilute \cite{lin2015} or in strongly correlated metals \cite{KADOWAKI}, which have both a small Fermi energy $E_F$, but not in dense and/or weakly correlated metals. The prefactor $A_2$, together with the electronic T-linear specific heat (i.e. the Sommerfeld coefficient, $\gamma$), allows us to put the system on the Kadowaki–Woods plot \cite{Rice1968,KADOWAKI,Wang2020}. Combined with $\gamma \simeq 20$ mJ.mol$^{-1}$.K$^{-2}$ \cite{yang2023charge, duan2021nodeless}, our data  yields: $A_2/\gamma^2 \simeq 3 \times 10^{-6}$ $\mu \Omega$.cm.mJ$^{-1}.$mol$^2$.K$^2$ in CsV$_3$Sb$_5$. This value is three times lower than the  Kadowaki-Woods ratio for strongly correlated metals $A_2/\gamma^2 [KW] = 10^{-5}$ $\mu \Omega$.cm.mJ$^{-1}.$mol$^2$.K$^2$ \cite{KADOWAKI}, but eight times larger than the Rice ratio for weakly correlated metals $A_2/\gamma^2 [Rice] =0.4 \times 10^{-6} \mu \Omega$.cm.mJ$^{-1}.$mol$^2$.K$^2$ \cite{Rice1968}. Thus, like many other Fermi liquids lying between these two fuzzy boundaries \cite{Wang2020}, CsV$_3$Sb$_5$ is a moderately correlated Fermi liquid . 

As for electron-phonon scattering, it is noteworthy that the amplitude of $A_5$ prefactor in CsV$_3$Sb$_5$ is large. In gold, with a comparable Debye temperature, $\Theta_D$, the tabulated resistivity data \cite{Matula} yields a fifty times smaller  T$^5$ prefactor ($A_5 (Au) \simeq 4 \times 10^{-9}\mu\Omega$.cm.K$^{-5}$). The unusually narrow window of T-square resistivity is connected to this feature. A relatively small $T^2$ term together with a relatively large amplitude of the $T^5$ term, confine the former to below 10 K.  

Let us now turn our attention to thermal transport. We measured the thermal conductivity ($\kappa$) of the same CsV$_3$Sb$_5$ sample using a one-heater–two-thermometers method using the same electrodes as for measurements of the electric resistivity.  Figure \ref{fig: Kappa} displays the thermal conductivity, $\kappa$,  in zero and finite magnetic field.  In the whole range of temperature (from 2 K to 30 K) and magnetic field ($B \leq 12$T), the amplitude of $\kappa$ is steadily diminished with increasing field. Our data  is  consistent with two other previous reports \cite{zhou2022anomalous,yang2023charge}, but not with two others \cite{hossain2025unconventional,kountz2024thermal}. The data sets are compared and contrasted in the appendix. As we will see below, the quantitative accuracy of our thermal transport data is backed by the recovery of the Wiedemann-Franz law.

Phonons and mobile electrons can both carry heat. In order to separate the phononic, $\kappa_{ph}$, and the electronic,  $\kappa_e$, components of $\kappa$, one can exploit the fact that, because of the electronic magnetoresistance, $\kappa_e$ is significantly affected by the magnetic field. In contrast, $\kappa_{ph}$ has no field dependence (at least, not a significant one). Such a procedure of using a magnetic field to separate $\kappa_{ph}$ and $\kappa_e$ has been used in semimetals \cite{white1958,jaoui2021thermal,jaoui2022, gourgout2024electronic,xie2024purity} as well as in metallic oxides \cite{jiang2023t,ling2025t}. 

 Specifically, we are interested to know the amplitude and the temperature dependence of the electronic Lorenz number, defined as  $L_e=\frac{\kappa_e}{T\sigma}$. Let us assume that magnetic field affects $\kappa$ and $\sigma$ in a similar way. In other words, at a given temperature, $L_e$ is independent of magnetic field. This is a reasonable assumption provided that we limit ourselves to small magnetic fields. This assumption allows us to quantify  $L_e$ experimentally by measuring $\Delta\kappa$ and $\Delta\sigma$, the field-induced variation of thermal and electrical conductivities and deducing  $L_e=\frac{\Delta\kappa}{T\Delta\sigma}$. These assumptions lead to the following expression for $\kappa_e$:

\begin{equation}
    \kappa_e=\sigma \frac{\Delta\kappa}{\Delta\sigma} 
\end{equation}

Afterwards, the phononic component can be obtained by a simple subtraction: $\kappa_{ph}=\kappa-\kappa_e$.

\begin{figure}[ht]
    \centering
     \begin{subfigure}{0.4\textwidth}
        \captionsetup{justification=raggedright,singlelinecheck=false}
        \caption{}
        \includegraphics[width=1\textwidth]{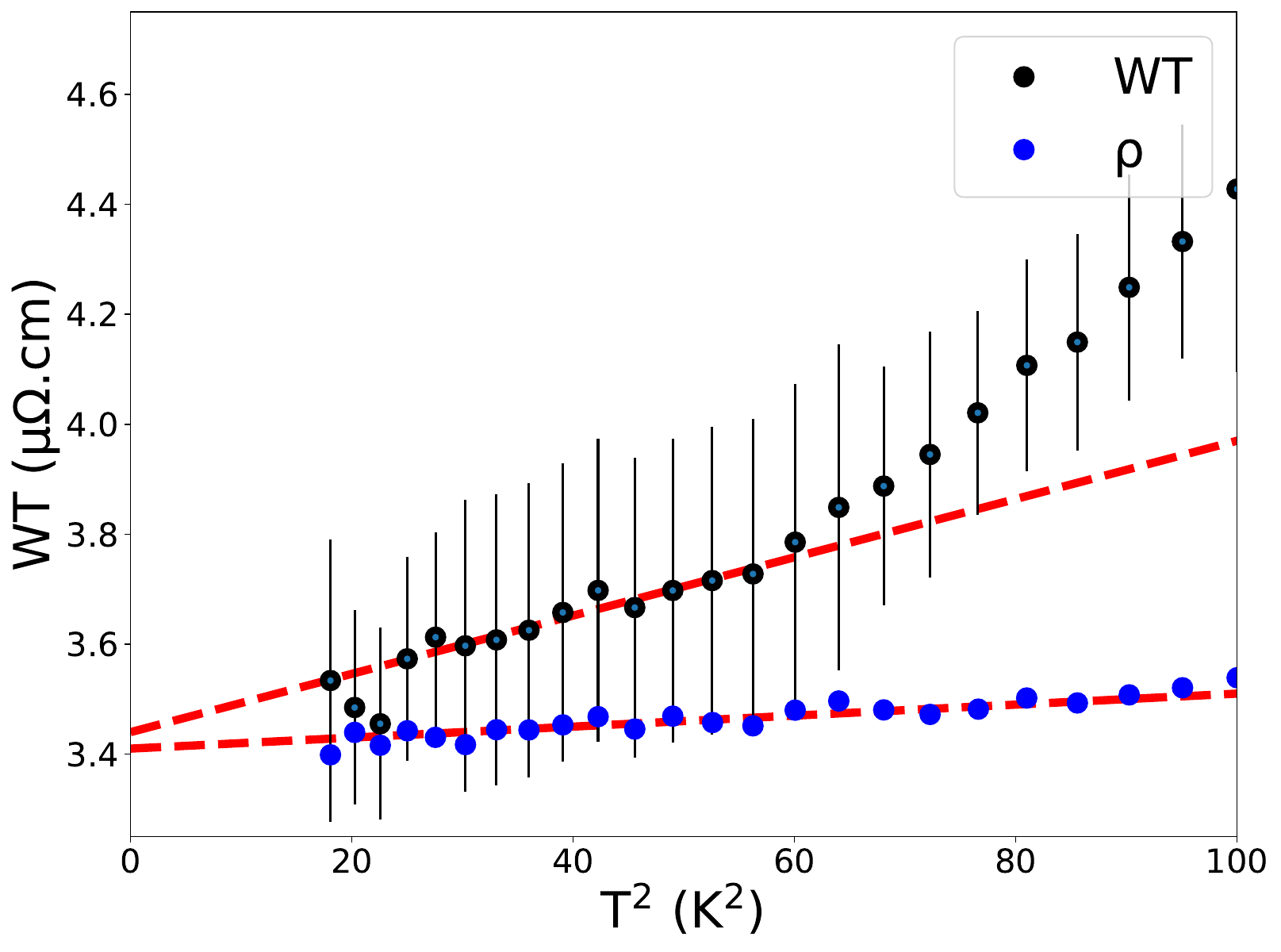}
     \end{subfigure}
     \begin{subfigure}{0.4\textwidth}
        \captionsetup{justification=raggedright,singlelinecheck=false}
        \caption{}
        \includegraphics[width=1\textwidth]{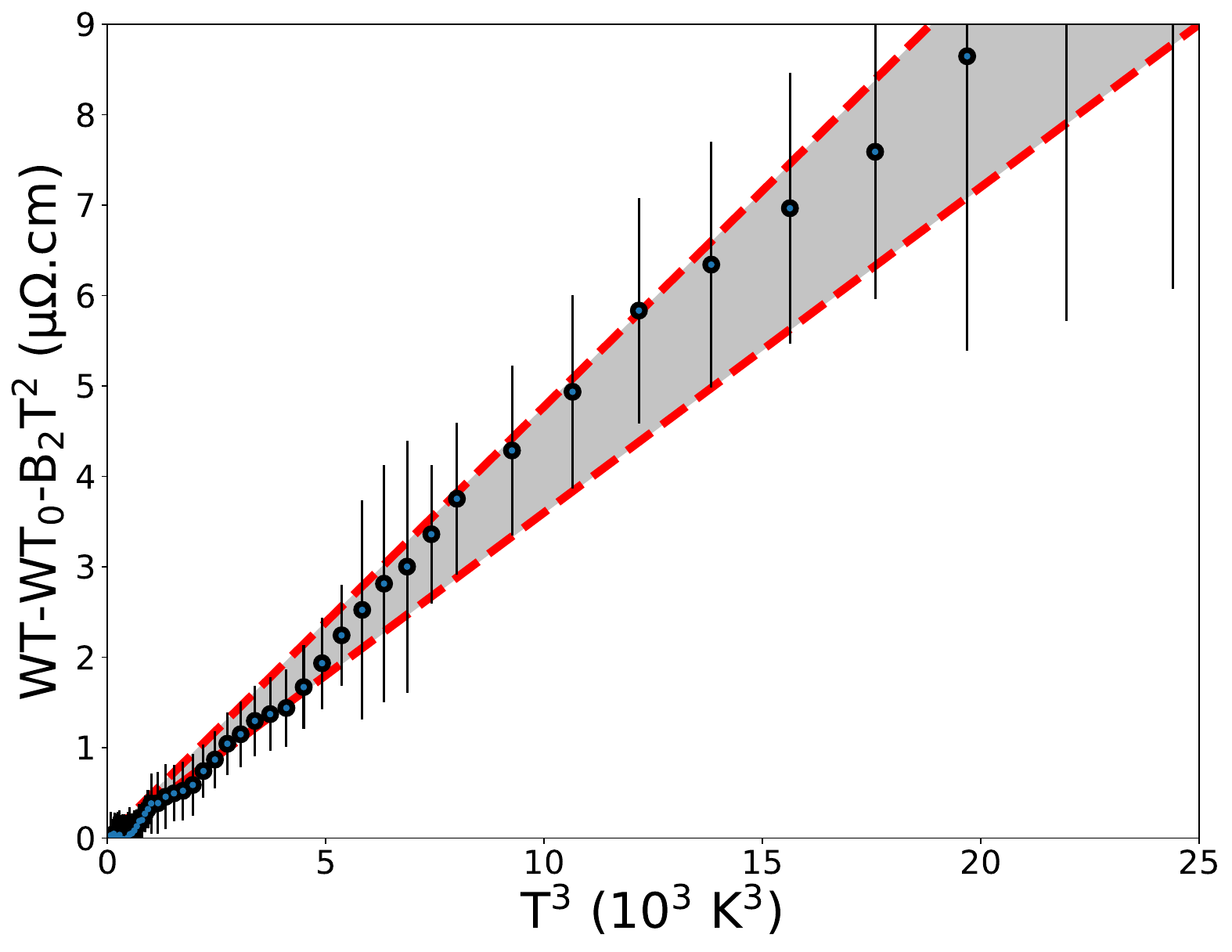}
     \end{subfigure}
     \caption{\justifying{Electrical thermal conductivity ($WT$). (a) WT at 0 T. (a) WT and $\rho$ versus temperature square. Red lines show T-square resistivity, for temperatures up to 7 K for $WT$ and 10 K for $\rho$. (b) WT without the constant and the quadratic term as function of temperature cube. The red lines show estimated maximum and minimum T-cube dependence for temperatures. The gray area shows the incertitude on the T-cube resistivity.}}
    \label{fig: WT}
\end{figure}

The results are depicted in Figure \ref{fig: separation}. Let us first consider the fate of the Wiedemann-Franz (WF) law, according to which, in the zero temperature limit, $L_e=L_0$ ($L_0=\frac{\pi^2}{3}\left(\frac{k_B}{e}\right)^2$ is the so-called Sommerfeld value). Figure \ref{fig: separation}a is a plot comparing the amplitude and temperature dependence of $\Delta\kappa/T$ and $L_0\Delta\sigma$, where $\Delta \kappa= \kappa (0T)- \kappa (1T)$  and $\Delta \sigma= \sigma (0T)- \sigma (1T)$. Between 4 K and 7 K, the two curves are close to each other as expected by the WF law. Figure \ref{fig: separation}b, which is a plot of the temperature dependence of $L_e/L_0$ is more explicit. Below 7 K, Wiedemann-Franz law is respected within experimental resolution. A downward departure is clearly visible. Such a deviation was previously observed in other metals, both strongly correlated such as CeRhIn$_5$ \cite{paglione2005heat} and UPt$_3$ \cite{Lussier1994} or in weakly correlated ones, such as  W \cite{Wagner1971} and WP$_2$ \cite{jaoui2018departure}.

\begin{figure*}[ht]
 \centering
\includegraphics[width=1\textwidth]{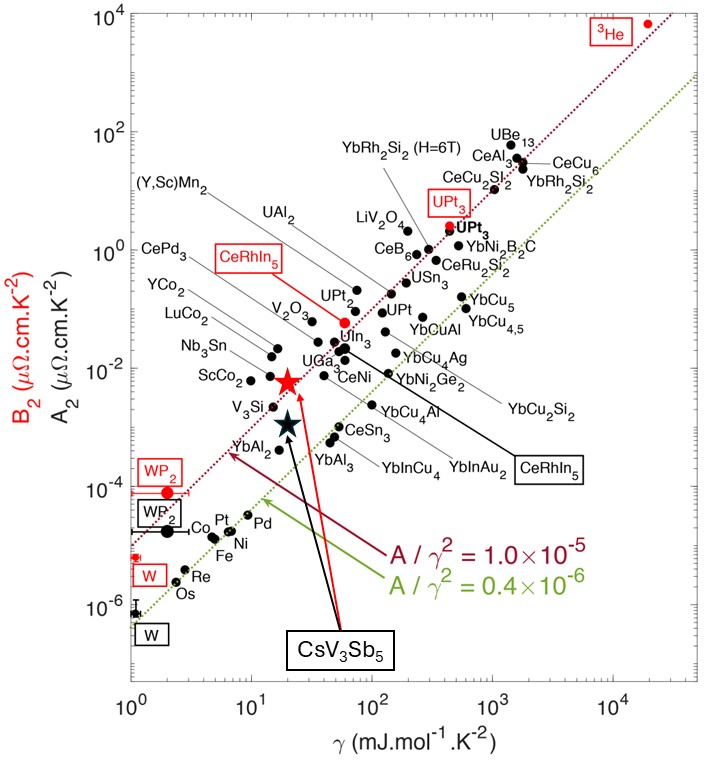}
\caption{\justifying{A Kadowaki-Woods plot, extracted from ref. \cite{jaoui2018departure}, with $A_2$ and $B_2$ data points for CsV$_3$Sb$_5$ included.} }
    \label{fig: KW}
\end{figure*}

Having quantified $L_e$, we can separate the electronic and the lattice components of the thermal conductivity. $\kappa_{ph}$ and   $\kappa_e$ are plotted in Figure \ref{fig: separation}c.  In the whole temperature range, $\kappa_{ph}$ is larger than $\kappa_e$. In the kinetic theory of gases \cite{kittel2018introduction}, an expression links  $\kappa_{ph}$ to the phonon mean free path, $l_{ph}$: $\kappa_{ph}=\frac{1}{3}C_{ph}v_{ph}l_{ph}$.  Combined with the reported amplitudes of the lattice specific heat, $C_{ph}$ \cite{yang2023charge}, and the sound velocity, $v_{ph}$ \cite{pang2023glasslike}, our data yields  $l_{ph}$. The result is shown in figure \ref{fig: separation}d. At T $\approx$ 4 K, $l_{ph}\simeq4$ $\mu$m  well below the sample thickness  (200 $\mu$m). Presumably, in CsV$_3$Sb$_5$, scattering of phonons by electrons impedes to reach the ballistic regime at above the superconducting transition. 

\begin{figure*}[ht]
 \centering
\includegraphics[width=0.7\textwidth]{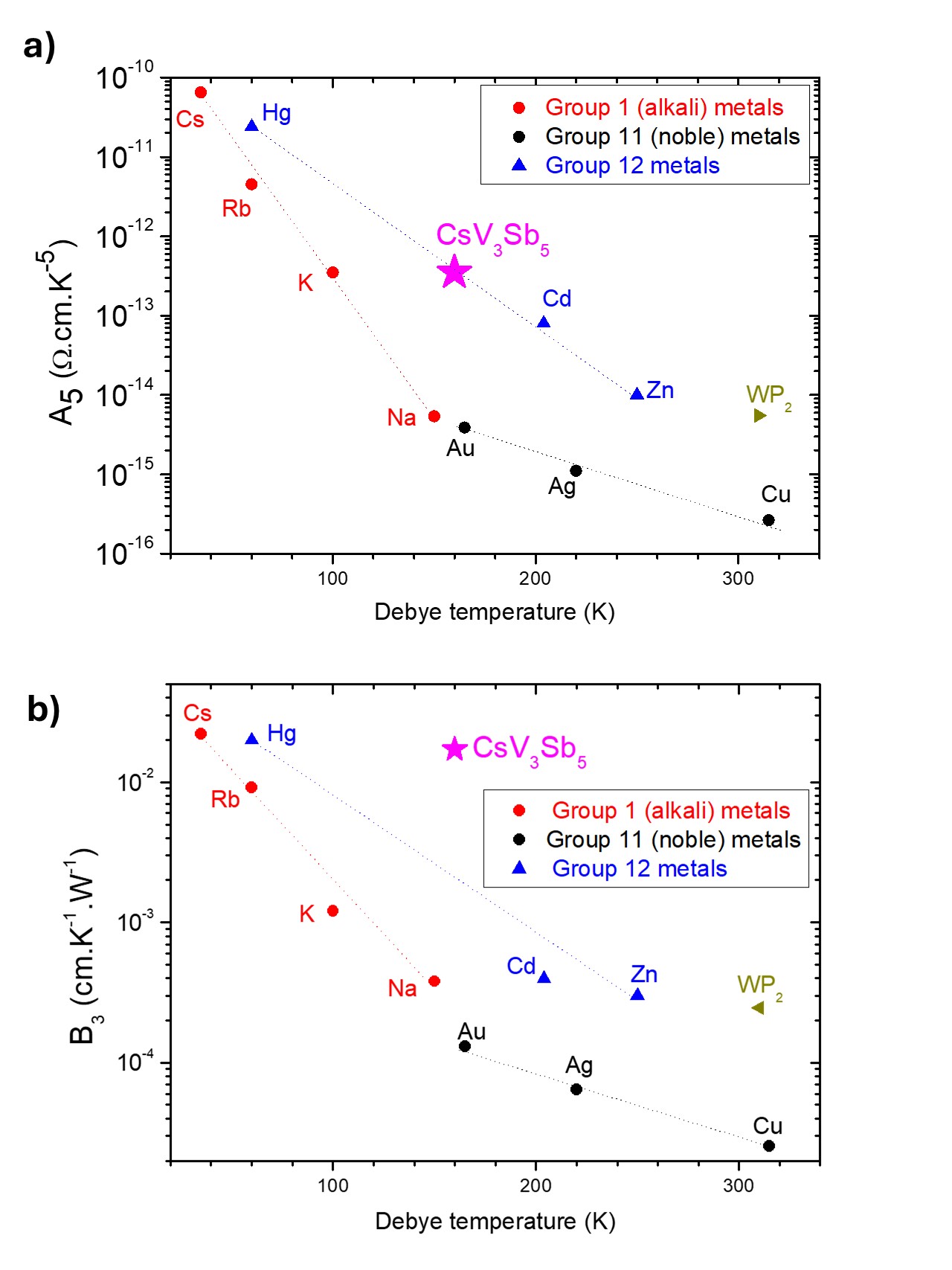}
\caption{\justifying{a) Amplitude of the prefactor of T$^5$ resistivity in CsV$_3$Sb$_5$ and in several elemental metals as a function of Debye temperature. Dashed lines indicate correlation between metals belonging to the same column of the periodic table. b) Same for the prefactor of thermal resistivity (expressed in its original units of cm.K$^{-1}$. W$^{-1}$  and without multiplication by $L_0$.) The main source of listed values for elements are two tables in ref.\cite{Klemens1956}. The amplitude of $B_3$ in gold is misprinted and the correct value is recovered by referring to  data reported in ref. \cite{White_1953}. We  used ref. \cite{Desai1984,RYAN1980158} for $A_5$ values of column 12 elements.} Dotted lines are guides for eye connecting metals belonging to the same group. The figure indicates that, considering its Debye temperature, CsV$_3$Sb$_5$ has relatively large e-ph prefactors.}
    \label{fig: A5B3}
\end{figure*}

Defining the electronic thermal resistivity as $WT=\frac{L_0T}{\kappa_e}$ allows a quantitative comparison between the electronic  heat transport and its charge counterpart. One expects the thermal resistivity to follow:

\begin{equation}
    WT=WT_0+B_2T^2+B_3T^3
    \label{thermal}
\end{equation}

The first term, $WT_0$  represents scattering by impurities and defects. The quadratic term is due to electron-electron scattering and finally, the cubic term is due to electron-phonon scattering. The validity of the WF law at T=0  imposes  $WT_0 =\rho_0$. Its invalidity at finite temperature implies that $B_2\neq A_2$. In the case of electron-phonon scattering, even the exponent changes. In thermal transport, it is $\propto T^3$, because only the variation of the phonon population matters. In electric transport, it is $\propto T^5$, because not all phonons matter equally.  Note that the prefactor of $T^5$ electric resistivity and the prefactor of $T^3$ thermal resistivity both scale inversely with the Debye temperature, but also with the specific material-dependent parameters, which set the strength of e-ph coupling in a given metal \cite{ziman2001electrons}.  

Figures \ref{fig: WT}a-b compare our data with what is expected by Equation \ref{thermal}. Figure \ref{fig: WT}a is a plot of $WT$ as a function of $T^2$. In the lowest temperature range, $WT$ is a linear in $T^2$ with an intercept, which yields $WT_0 =\rho_0$.  The slope allows the quantification of $B_2$, significantly larger than $A_2$, which sets the slope of electric resistivity. Above 8 K,  there is an upward deviation pointing to the emergence of a higher exponent. Plotting $WT-WT_0-B_2T^2$ as a function of $T^3$ in Fig. \ref{fig: WT}b, allows to identify a cubic regime below $\approx$ 20 K. 

The amplitude of $B_2$, extracted by a quadratic fit to $WT$ yields $B_2=6.6\pm0.9$ n$\Omega$.cm.K$^{-2}$. The $B_2/A_2$ ratio in CsV$_3$Sb$_5$ is 5.2.  This ratio, which is found to exceed unity in all explored cases, ranges between 1.6 and 7 (See table \ref{tab:A2 B2} in the appendix). Our result, by putting CsV$_3$Sb$_5$ in the upper part of this range, indicates that a large $B_2/A_2$ can be attained even in the absence of compensation, theoretically proposed \cite{li2018} as a possible driver of a large  $B_2/A_2$ ratio, a feature assumed to be forbidden by an early theoretical study \cite{Herring1967}. Figure \ref{fig: KW} shows how $B_2$ and $A_2$ of CsV$_3$Sb$_5$ compares to that of other Fermi liquids in a Kadowaki-Woods plot.

We saw above that the remarkably large amplitude of $A_5$ points to a large collision cross section between electrons and phonons in CsV$_3$Sb$_5$. This is confirmed in the thermal transport channel. The relative amplitude of $B_3$, compared to other metals with a similar Debye temperature, is even larger. 

Figure \ref{fig: A5B3} compares the amplitude of $A_5$ and $B_3$ in CsV$_3$Sb$_5$ with a number of elemental metals. In each family, one can see a specific correlation between  $A_5$ and $\Theta_D$. A similar (albeit less straightforward) family-dependent correlation between $B_3$ and $\Theta_D$ is also visible. Note that group 12 elements (Zn,Cd, Hg) with a HCP structure have a larger $A_5$ and $B_3$ than group 11 elements (Cu,Ag, Au) reflecting a larger electron-phonon cross section due to the proximity of the structural transition occurring between the two columns \cite{Subedi_2025}. The absence of superconductivity in noble and alkali metals is often attributed to the weakness of e-ph coupling. Group 12 elements have superconducting ground states and known to have a larger e-ph coupling constant. 

\section{Discussion}
In absence of a systematic study of other members of  AV$_3$Sb$_5$, we can limit ourselves to a simple observation. The e-ph collision cross section in CsV$_3$Sb$_5$, normalized to its Debye temperature is larger than these elemental metals. Note that this statement is about the CDW state. An estimation of the e-ph coupling constant from T-linear resistivity (see the appendix) finds that $\lambda \simeq 0.7$, in agreement with a previous study \cite{Zhong2023}. 

Thus, our results indicate that in CsV$_3$Sb$_5$, the e-ph is significantly enhanced. This is consistent with what was previously found by studies employing neutron scattering \cite{Xie2022}, Raman spectroscopy \cite{He2024} and angle-resolved photoemission spectroscopy \cite{Zhong2023}. We note  that  \textit{Ab Initio} calculations  \cite{Subedi2022} find a close competition between crystal structures with different symmetries to become the ground state of this solid. The consequences of this competition for e-ph collision rate emerges as a subject for future theoretical studies. 

In summary, by measuring electrical and thermal conductivities of CsV$_3$Sb$_5$, we separated the electronic and the phononic contributions to heat transport. We found that the transport properties can be depicted in the standard picture of transport in metallic Fermi liquids. Comparison with other metallic Fermi liquids indicates that electrons suffer a moderate rate of collisions with other electrons and a strong one with phonons.

\section{Acknowledgments}
This work was supported by the Agence Nationale de la Recherche (ANR-25-CE30-5730-01) and by a grant attributed by the Ile de France regional council. SDW and ACS gratefully acknowledge support via the UC Santa Barbara NSF Quantum Foundry funded via the Q-AMASE-i program under award DMR-1906325.

\section{Data Availability}
The research data is available upon request sent to the authors.
\clearpage
\begin{center}{\large\bf APPENDIX}\\
\end{center}

\renewcommand{\thesection}{A\arabic{section}}
\renewcommand{\thetable}{A\arabic{table}}
\renewcommand{\thefigure}{A\arabic{figure}}
\renewcommand{\theequation}{A\arabic{equation}}

\setcounter{section}{0}
\setcounter{figure}{0}
\setcounter{table}{0}
\setcounter{equation}{0}

\section{Superconducting transition and electric resistivity in a magnetic field.}
\begin{figure}[ht]
     \begin{subfigure}{0.5\textwidth}
        \captionsetup{justification=raggedright,singlelinecheck=false}
        \caption{}
        \includegraphics[width=1\textwidth]{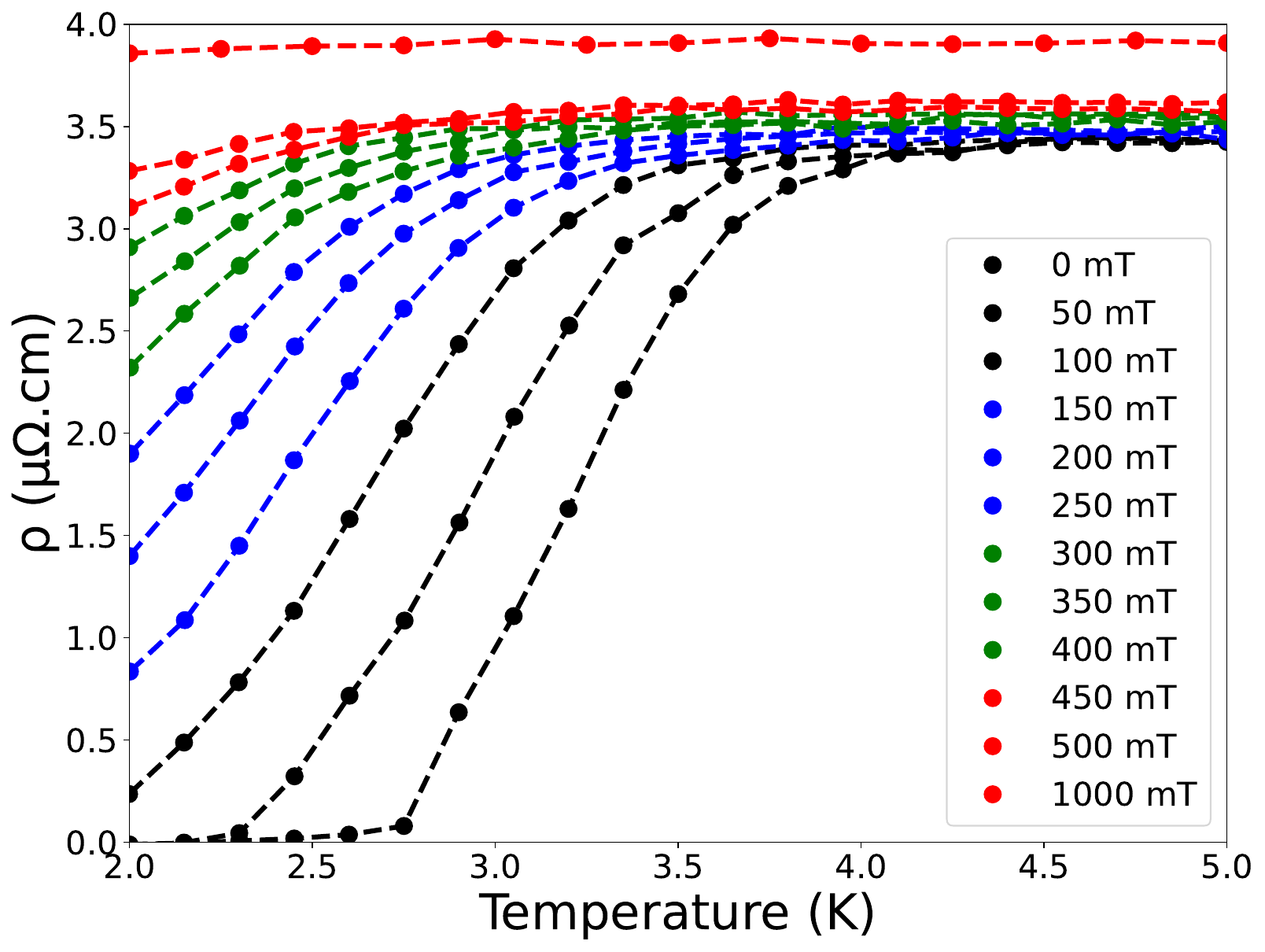}
     \end{subfigure}
     \begin{subfigure}{0.49\textwidth}
        \captionsetup{justification=raggedright,singlelinecheck=false}
        \caption{}
        \includegraphics[width=1\textwidth]{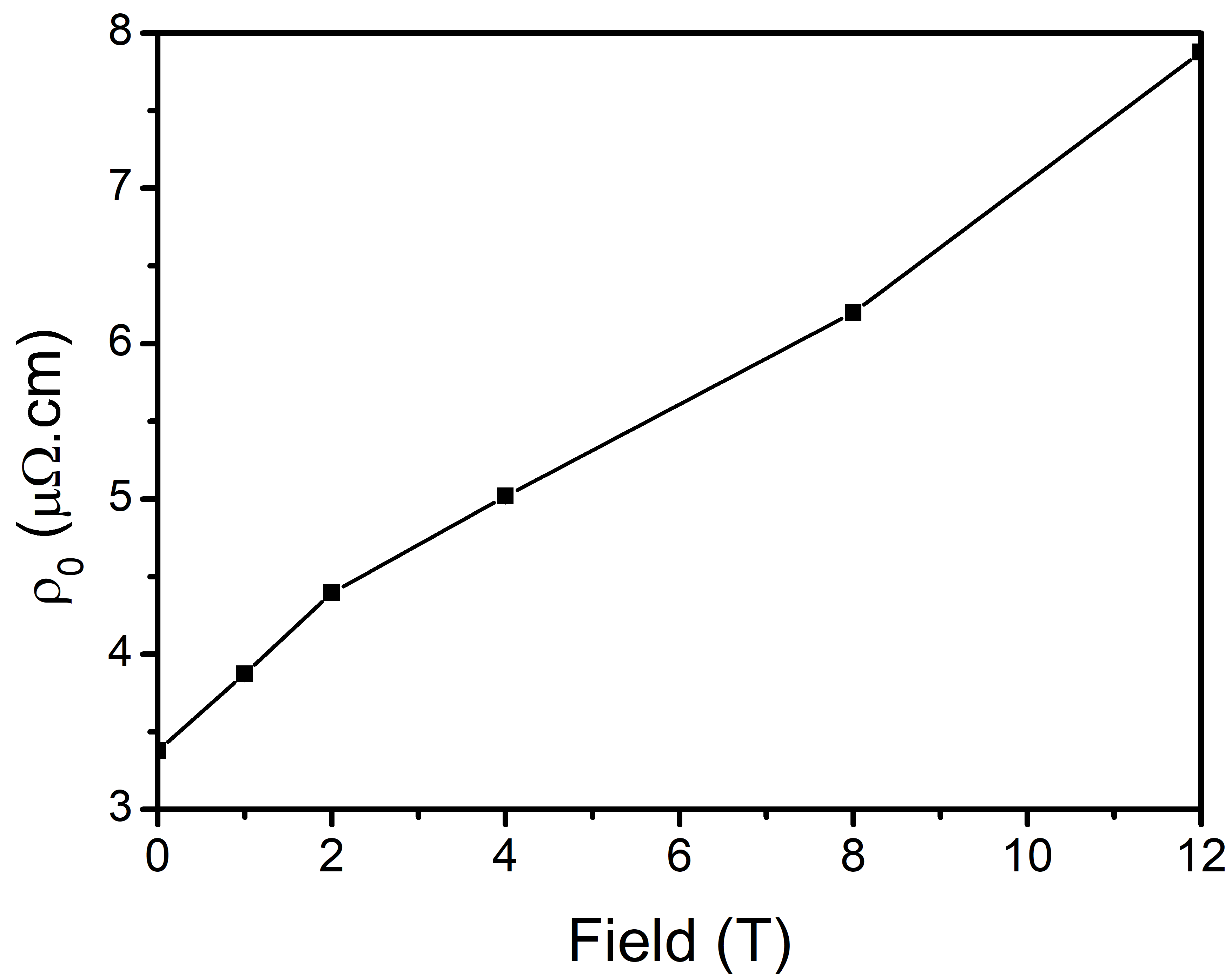}
     \end{subfigure}
     \caption{\justifying{(a) In-plane resistivity of CsV$_3$Sb$_5$ between 2 K and 5 K with applied magnetic field up to 1 T. (b) Residual resistivity as function of magnetic field.}}
    \label{fig: SM rho}
\end{figure}

Figure \ref{fig: SM rho}a displays electrical resistivity between 2 K and 5 K for small applied magnetic field such that superconducting transition is visible. At zero magnetic field, the superconducting transition starts at 4.25 K. Figure \ref{fig: SM rho}b shows the residual resistivity ($\rho_0$) as a function of magnetic field. $\rho_0$ increases  roughly  linearly with magnetic field.

\section{High-temperature electrical resistivity and e-ph coupling}

\begin{figure}[ht]
    \centering
        \includegraphics[width=0.48\textwidth]{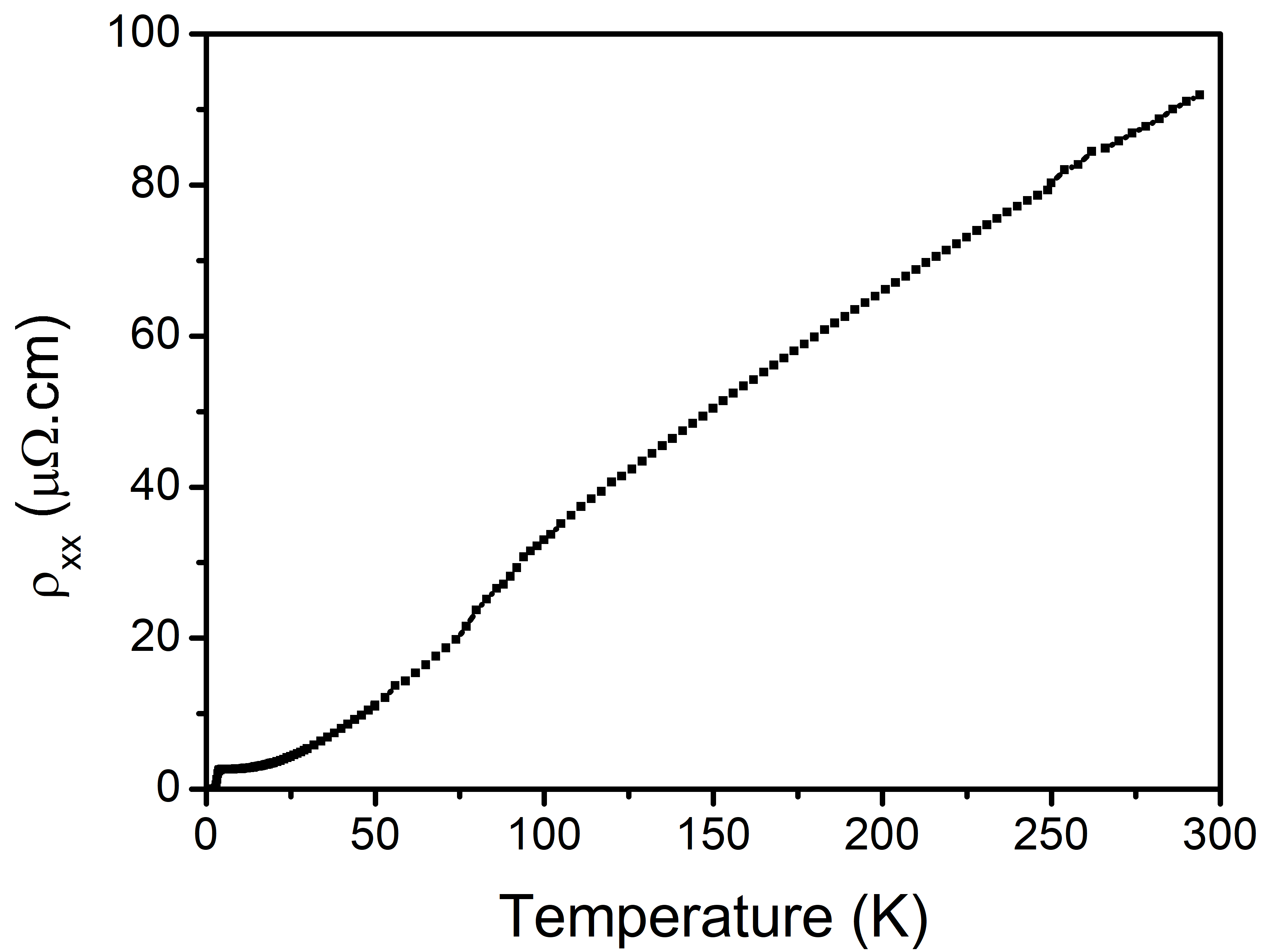}
   \caption{\justifying{In plane resistivity of CsV$_3$Sb$_5$ as function of temperature.} } 
    \label{fig: SM RT}
\end{figure}

Figure \ref{fig: SM RT} is a plot of the in-plane resistivity ($\rho_{xx}$) of CsV$_3$Sb$_5$ from 2 K to 294 K. The resistivity becomes linear in temperature when $T \geq 200 K$. with an estimated  slope of $\frac{d\rho}{dT} \simeq 0.25~\mu\Omega\cdot$cm·K$^{-1}$.

According to the Bloch–Grüneisen picture of electron–phonon scattering, the resistivity becomes T-linear once $T$ is comparable to or larger than the Debye temperature $\theta_D$. In this limit \cite{allen1987}:
\begin{equation}
\frac{d \rho}{dT}= 2\pi \lambda k_B \frac{\hbar}{e^2} \frac{n}{m^{*}}
\end{equation}

Here, $\lambda$ is the electron–phonon coupling constant,  $n$  the carrier density and m$^{*}$ the effective mass of the carriers. Above the CDW phase, the Fermi surface is formed by two large bands. According to quantum oscillation studies under pressure \cite{Zhang2024}, the largest band, labeled $\zeta$, has a frequency of $F_{\zeta} = 8.16$ kT, an effective mass of $m^{*} = 1.6m_0$, and a two-dimensional character. For this quasi-2D band, we estimate the carrier density to be $n \simeq 1.3 \times 10^{22}$ cm$^{-3}$, which leads to $\lambda = 0.7$, comparable to the intermediate value $\lambda = 0.45\text{–}0.6$ estimated by laser-based angle-resolved photoemission spectroscopy \cite{Zhong2023}.

\begin{figure}[ht]
    \centering
        \includegraphics[width=0.48\textwidth]{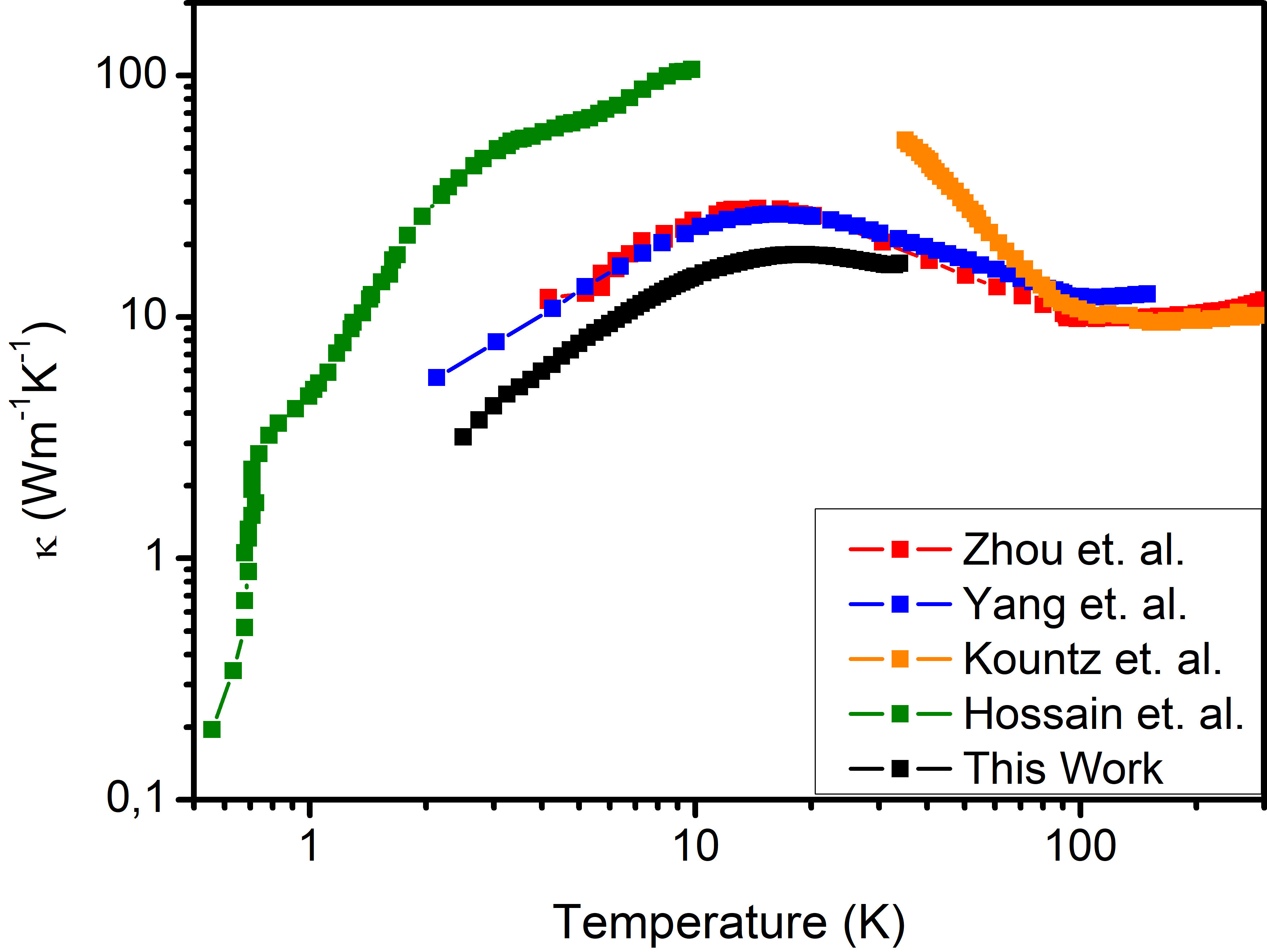}
   \caption{\justifying{Zero-field thermal conductivity of CsV$_3$Sb$_5$ reported by different groups \cite{zhou2022anomalous,kountz2024thermal,yang2023charge,hossain2025unconventional}. A large scattering is observed between five sets of data. Our data appears to agree with what was reported by Zhou \textit{et al.} \cite{zhou2022anomalous} and by Yang \textit{et al. }\cite{yang2023charge}. There is a significant difference between these three data sets and what was reported by Hossain \textit{et al.} \cite{hossain2025unconventional}and by Kountz \textit{et al.} \cite{kountz2024thermal}.} } 
    \label{fig: SM compar litt}
\end{figure}

\section{Comparison with previously reported data}
Thermal transport in  CsV$_3$Sb$_5$ has been studied by several groups \cite{zhou2022anomalous,kountz2024thermal,yang2023charge,hossain2025unconventional, zhao2024ultralow,zhang2024large}. 

As seen in Fig. \ref{fig: SM compar litt}, when the temperature range of interest overlaps, there is a significant discrepancy between some sets of data. One can see that between 2 K and 10 K, our data is in agreement with what was reported by two other groups, but almost 8 times lower than what was reported by another. Moreover, at 30 K, there is a significant discrepancy between what is reported by these two groups and what was reported by another group. However, this discrepancy vanishes at 100 K, that is above the CDW transition temperature. 

All groups used an identical method. However, they did not systemically report on the verification of their thermal transport data by recovering the Wiedemann-Franz law in the zero-temperature limit. Nevertheless, it is unlikely that this discrepancy is due to a measurement error. A more interesting possibility is that small differences between single crystalline samples drastically affects the increase in the thermal conductivity induced by the CDW transition.  This hypothesis is to be tested by future studies covering the whole temperature range.

At this stage, it is fair to make a few qualitative remarks.   Two different kinds of behavior are visible in Fig. \ref{fig: SM compar litt} .  Three sets of data almost overlap. This is the case of present study and those reported by Zhou \textit{et al.} \cite{zhou2022anomalous} and by Yang \textit{et al. }\cite{yang2023charge}). On the other hand, and in contrast with this consistency, the data reported by Hossain \textit{et al.} \cite{hossain2025unconventional} and restricted to below 10K, is almost five times larger. Kountz \textit{et al.} \cite{kountz2024thermal} also report a thermal conductivity three times larger than the the data reported by by Zhou \textit{et al.} \cite{zhou2022anomalous} and Yang \textit{et al. }\cite{yang2023charge} between 40 K and 80 K. Because of the difference in the temperature ranges explored , the consistency between \cite{hossain2025unconventional} and \cite{kountz2024thermal} can only be guessed. Since the remarkably large thermal resistivity driven by  e-ph coupling is expected to be strain-dependent,  the presence of uncontrolled strain is a plausible source of this sample dependence.

\section{Thermal conductivity at magnetic fields exceeding 1T}
\begin{figure}[ht]
    \centering
     \begin{subfigure}{0.49\textwidth}
        \captionsetup{justification=raggedright,singlelinecheck=false}
        \caption{}
        \includegraphics[width=1\textwidth]{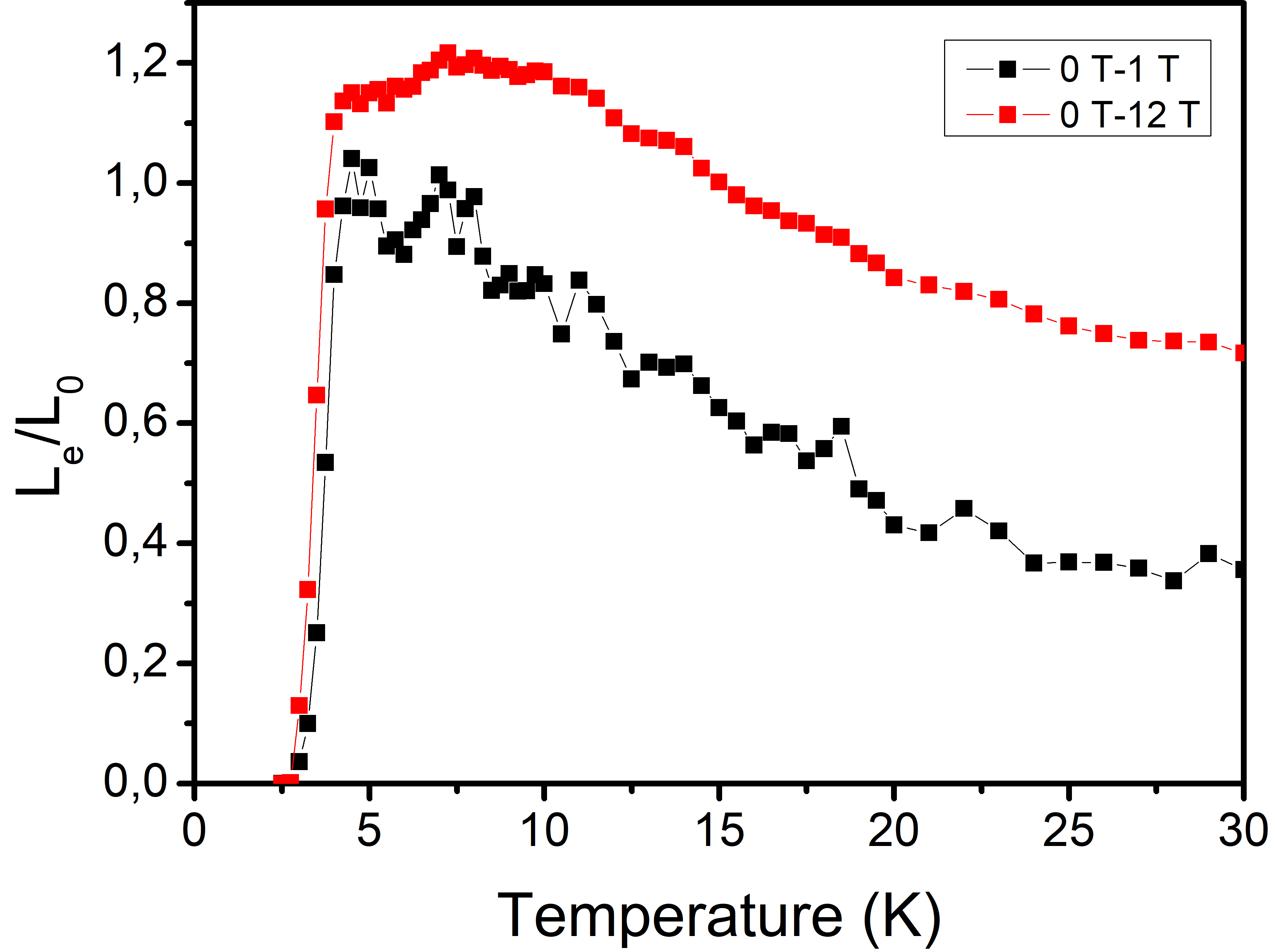}
     \end{subfigure}
     \begin{subfigure}{0.49\textwidth}
        \captionsetup{justification=raggedright,singlelinecheck=false}
        \caption{}
        \includegraphics[width=1\textwidth]{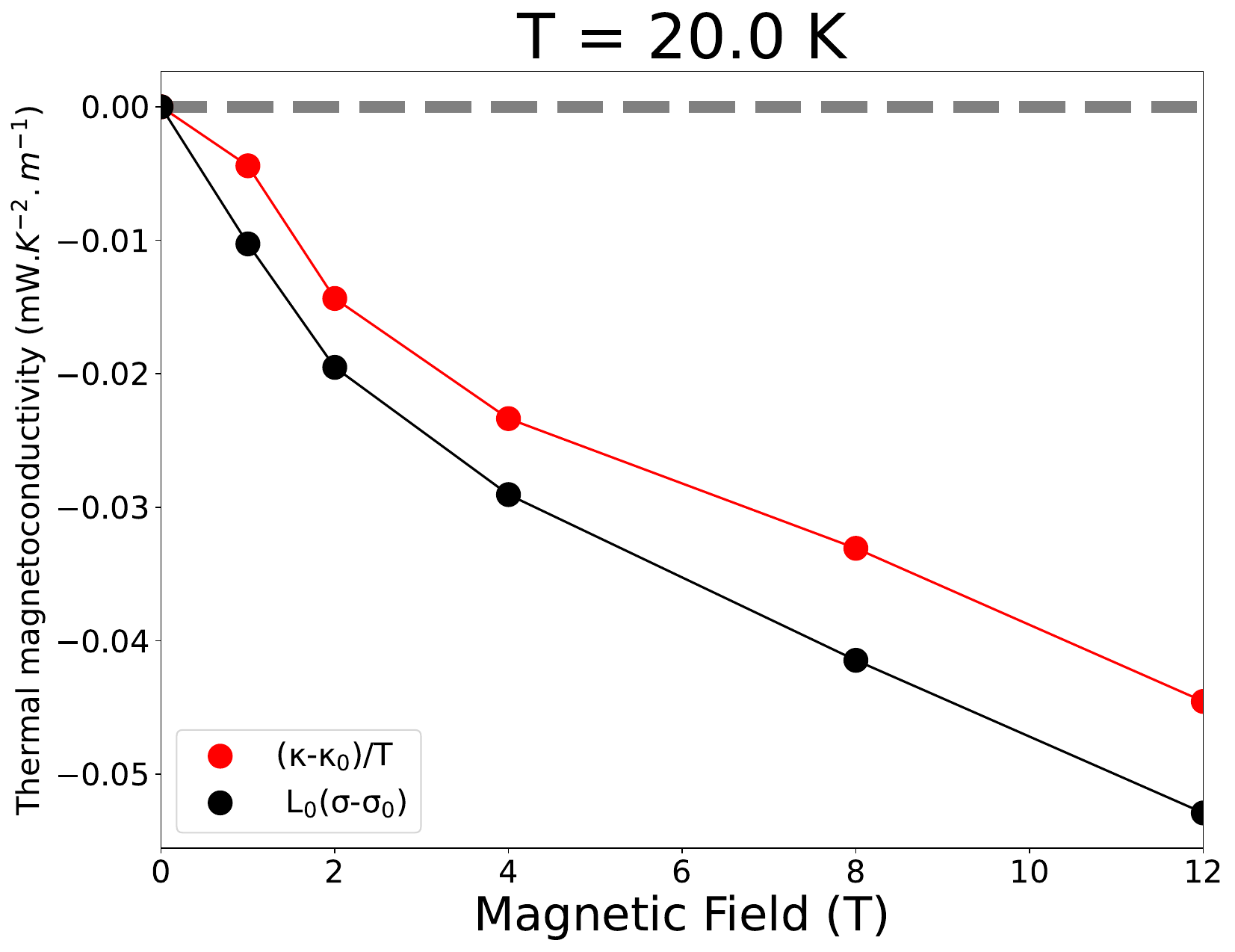}
     \end{subfigure}
     \caption{\justifying{(a) $L_e/L_0$ calculated from the difference in conductivities between 0 T and 1 T or 0 T and 12 T. (b) Difference of conductivities between zero magnetic field ($\sigma_0$ and $\kappa_0$) and finite, at 20 K.}}
    \label{fig: SM L0}
\end{figure}

Figure \ref{fig: SM L0}a plots estimated $L_e/L_0$ using the difference in conductivities between 0 T and 1 T (black) or 0 T and 12 T (red). In both cases, $L_e/L_0$ tends to zero below 4 K due to the apparition of superconductivity, in this case Wiedemann-Franz law does not apply. For temperatures above 4 K, difference between 0 T and 12 T yields to a higher $L_e$ compares to the other case. In particular, $L_e/L_0$ seems to tend to 1.2 at low temperature. This overall increase of $L_e$ is not only restricted to conductivities at 12 T but is also observed when we subtracted conductivity between 0 T and any field above 1 T. This indicates that  when the magnetic field when exceeds 1 T, either $L_e$ or the phonon thermal conduction are significantly affected by the magnetic field field.

Figure \ref{fig: SM L0}b plots the difference between zero-field and finite-field thermal and electric conductivities, at 20 K. For all magnetic fields, the difference in electrical conductivity is larger than its thermal counterpart, confirming that  $L_e/L_0 <1$. A kink between 0 T and 1 T is visible in the field dependence of thermal conductivity, but not in the the field dependence of electric conductivity. Clarifying the origin of this field-induced change in thermal transport is a task for future studies.

\section{The T-square electric and thermal resistivity prefactors}
Table \ref{tab:A2 B2} lists a selection of reported amplitudes of $A_2$ and $B_2$,  the prefactors of T-square electric and thermal resistivities. Both are expressed in units of n$\Omega.$cm.K$^{-2}$.   In all cases, $B_2/A_2 >1$. CsV$_3$Sb$_5$ does not differ from other cases. Taken at its face value, this remarkably large  $B_2/A_2$ indicates that, in this solid, a large fraction of e-e scattering is either small angle or Normal (i.e. not Umklapp). 

The differentiation between ``vertical” and ``horizontal” scattering events pulls down the prefactor for electrical resistivity. Herring \cite{Herring1967} argued that this difference is bounded  an therefore the $B_2/A_2 $ cannot exceed 2.  Li and Maslov \cite{li2018} argued  that in a compensated semimetal this boundary can be exceeded. The fact that CsV$_3$Sb$_5$ is not a compensated metal implies that compensation is not a necessary condition for $B_2/A_2 > 2$. 

\begin{table}[ht]
    \centering
    \begin{tabular}{|c|c|c|c|c|}
    \hline
        Material & $A_2$ & $B_2$ & $B_2/A_2$& Ref.\\
        \hline
        W &$8.7\times 10^{-4}$&$6 \times 10^{-3}$&6.9&\cite{Wagner1971}\\
        Bi &12 &35 &2.9 &\cite{gourgout2024electronic}\\
        Sb & 0.3& 0.6&2.0 &\cite{jaoui2021thermal}\\
       WTe$_2$ & 4.5& 11&2.4 &\cite{xie2024purity}\\
       WP$_2$ & 0.017&0.074 &4.6 &\cite{jaoui2018departure}\\
       CeRhIn$_5$ &21 & 57& 2.7& \cite{paglione2005heat}\\
       RuO$_2$ &0.071 & 0.26& 3.7& \cite{ling2025t}\\
        CsV$_3$Sb$_5$ &1.26 & 6.6& 5.2& This work\\
         \hline
    \end{tabular}
    \caption{\justifying{Reported values of electric and thermal T-square prefactors.}}
    \label{tab:A2 B2}
\end{table}

\bibliography{sample}

\end{document}